\documentclass[aps,prc,twocolumn,superscriptaddress,showpacs]{revtex4-1}

\usepackage[dvips]{graphicx}
\usepackage{amsmath,amssymb}
\usepackage[T1]{fontenc}

\begin{document}

\title{Thermal quasiparticle random-phase approximation with  Skyrme interactions and
supernova neutral-current neutrino-nucleus reactions}

\author{Alan~A.~Dzhioev}
\email{dzhioev@theor.jinr.ru}
\affiliation{Bogoliubov Laboratory of Theoretical Physics, JINR, 141980, Dubna, Russia}
\author{A.~I.~Vdovin}
\email{vdovin@theor.jinr.ru}
\affiliation{Bogoliubov Laboratory of Theoretical Physics, JINR, 141980, Dubna, Russia}
\author{G.~Mart\'{\i}nez-Pinedo}
\email{gabriel.martinez@physik.tu-darmstadt.de}
\affiliation{Institut f{\"u}r Kernphysik, Technische Universit{\"a}t Darmstadt, 64289 Darmstadt, Germany}
\affiliation{GSI Helmholtzzentrum f\"ur Schwerionenforschung, Planckstr. 1, 64291 Darmstadt, Germany}
 \author{J.~Wambach}
 \email{jochen.wambach@physik.tu-darmstadt.de}
\affiliation{Institut f{\"u}r Kernphysik, Technische Universit{\"a}t Darmstadt, 64289 Darmstadt, Germany}
\affiliation{GSI Helmholtzzentrum f\"ur Schwerionenforschung, Planckstr. 1, 64291 Darmstadt, Germany}
\author{Ch.~Stoyanov}
\email{stoyanov@inrne.bas.bg}
\affiliation{Institute for Nuclear Research and Nuclear Energy, 
1784 Sofia, Bulgaria}

\date{\today}

\begin{abstract}
The Thermal Quasiparticle Random-Phase Approximation is combined with the Skyrme energy density functional method (Skyrme-TQRPA) to study the response of a hot nucleus to an external perturbation.  For the sample nuclei, $^{56}$Fe and $^{82}$Ge, the Skyrme-TQRPA is applied to analyze thermal effects on  the strength function of charge-neutral Gamow-Teller transitions which dominate neutrino-nucleus reactions at $E_\nu \lesssim 20$~MeV. For the relevant supernova temperatures we calculate the cross sections for inelastic neutrino scattering. We also apply the method to examine the rate of neutrino-antineutrino pair emission by hot nuclei. The cross sections and rates are compared with those obtained earlier  from the TQRPA calculations based on the phenomenological Quasiparticle-Phonon Model Hamiltonian. For inelastic neutrino scattering on $^{56}$Fe we also compare the Skyrme-TQRPA results to those obtained earlier from a  hybrid approach that combines shell-model and RPA
calculations.
\end{abstract}




\maketitle

\section{Introduction}

The significant role played by processes involving neutrinos in core-collapse supernovae is well
known~\cite{Janka_PhysRep442, Balasi_PPNPh85}. During the collapse, neutrinos carry away
energy helping to maintain a low entropy. As a result, nucleons reside
primarily in nuclei.  At densities of $\rho\gtrsim10^{11}\,\mathrm{g\,cm}^{-3}$ neutrino interactions with matter become important, leading to neutrino trapping and thermalization.  Moreover, neutrino energy deposition plays an important role in modeling supernova  explosions. Obviously, reliable supernova simulations require a detailed neutrino transport and
should in principle include all potentially important neutrino reactions with supernova matter.

Inelastic neutrino scattering on nuclei is a process which contributes to establishing equilibrium between neutrinos and matter. Up to moderate neutrino energies ($E_\nu\lesssim 20$~MeV), the scattering is dominated by charge-neutral Gamow-Teller (GT$_0$) transitions and therefore the GT$_0$ strength distribution is of special importance. At high supernova temperatures, one can expect the thermal population of nuclear excited states. The Gamow-Teller transitions from nuclear excited states  remove the reaction
threshold and dominate the  cross section at low neutrino energies~\cite{Fuller_APJ376}. Moreover, in the supernova environment GT$_0$ transitions determine the decay of thermally excited nuclei by neutrino-antineutrino pair emission. The astrophysical importance of this process was first realized by
Pontecorvo~\cite{Pontecorvo_PLB1}, who noticed that the emission of $\nu\bar\nu$ pairs by excited nuclei may be a powerful mechanism for the energy loss by stars.

Sampaio \textit{et al}. have studied thermal effects on  inelastic neutrino scattering on nuclei in supernova environment~\cite{Sampaio_PLetB529}. The study was
performed on a representative set of iron-group nuclei ($A\sim 60$), using results from large-scale shell-model calculations (LSSM) of the allowed GT$_0$ transitions.
It was shown that finite temperature increases the low-energy  ($E_\nu\mathbf{\lesssim}10$~MeV) cross section significantly and this effects is most pronounced for even-even nuclei (e.g., $^{56}$Fe).
In~\cite{Juodagalvis_NPA747}, the calculations were extended to higher neutrino energies within a hybrid approach that derives forbidden multipole contributions
from RPA calculations.

Although contemporary  shell-model calculations are capable of describing GT$_0$ distributions rather well, an explicit calculation of the cross section at finite temperature ($T\gtrsim  1$~ MeV) is computationally
unfeasible due to too many thermally populated states. To overcome this problem, Refs.~\cite{Sampaio_PLetB529, Juodagalvis_NPA747} apply the Brink hypothesis,
which states that strength distributions for excited states are the same as for the nuclear ground state, but shifted by excitation energy.
However, the validity of the Brink hypothesis for the GT strength function is not obvious and its violation is confirmed by the shell-model Monte Carlo studies at finite temperature~\cite{Radha_PRC56} and most recently by the shell-model calculations~\cite{Misch_PRC90}.
It should also be mentioned, that present computer capabilities allow large-scale shell-model calculation only for nuclei with $A\leqslant 65$,  whereas neutrino reactions with heavier mass and neutron-rich nuclei also may play an important role in core-collapse supernovae.

For the response of hot nuclei to an external perturbation in~\cite{Dzhioev_IJMPE18},  a so-called thermal quasiparticle random-phase approximation (TQRPA) was proposed in the framework of a statistical approach. The method is based on the thermo-field dynamics (TFD) formalism~\cite{Takahashi_IJMPB10,Umezawa1982,Ojima_AnPhys137} and enables the computation of a temperature-dependent strength function without assumption of Brink's hypothesis.
In Refs.~\cite{Dzhioev_PAN74, Dzhioev_PRC89,Dzhioev_PAN77},  the TQRPA was applied to compute cross sections and rates for neutral-current neutrino-nucleus reactions at the relevant supernova temperatures. It was shown that, although the TQRPA reveals the same thermal effects as reported in Refs.~\cite{Sampaio_PLetB529, Juodagalvis_NPA747}, the observed thermal enhancement of the cross sections turns out to be several times larger.  It was found that this discrepancy stems from two main reasons:  (i) Within the TQRPA,  temperature rise shifts the GT$_0$ centroid to lower energies and makes  possible low-energy GT$_0$ transitions which are Pauli blocked at $T=0$.  These effects are not present in calculations based on the Brink hypothesis.  (ii) Unlike the approach of Refs.~\cite{Sampaio_PLetB529, Juodagalvis_NPA747}, the principle of detailed balance is not violated within the TQRPA and it results in a larger strength for downward
GT$_0$ transitions from excited states. It is clear that both these factors should favour neutrino inelastic scattering.

In Refs.~\cite{Dzhioev_PAN74, Dzhioev_PRC89,Dzhioev_PAN77}, the TQRPA calculations were based on a Hamiltonian containing a Woods-Saxon potential supplemented by a pairing interaction and a schematic residual particle-hole interaction of a separable form. This Hamiltonian is usually referred to as the Hamiltonian of the quasiparticle-phonon model (QPM)~\cite{Soloviev1992}. The parameters of  the Hamiltonian are adjusted locally, i.e., to properties of a nucleus under consideration. In this paper, we extend our studies and perform self-consistent calculations for neutral-current neutrino-nucleus reaction combining the TQRPA approach with the Skyrme energy density functional. The calculations are performed within the finite-rank separable approach, which expands the Skyrme residual interaction into a sum of separable terms in a systematic manner~\cite{Giai_PRevC57, Sever_PRC66}. The factorization considerably reduces the computational effort of the TQRPA while maintaining high accuracy. This allows one to perform systematic studies over a wide range of nuclei including those far from stability.

The paper is organized as follows: In Sec.~\ref{formalism}, we briefly outline how to compute the thermal strength function within the TFD formalism. The method of separability of the Skyrme residual interaction and the TQRPA approach are summarized in Sec.~\ref{formalism}. The results of the numerical calculations are presented and discussed in Sec.~\ref{results} for the sample nuclei~$^{56}$Fe and $^{82}$Ge. The results are compared  with those obtained earlier from the TQRPA calculations based on the QPM Hamiltonian. For $^{56}$Fe we also compare the results to those obtained earlier from the hybrid approach~\cite{Juodagalvis_NPA747}, which combines shell-model and RPA calculations. In Sec.~\ref{conclusion}, we draw conclusions and give an outlook for future studies. The derivation of the TQRPA equations for the finite-rank separable Skyrme interaction is given in the Appendix.

\section{Formalism}\label{formalism}

\subsection{Thermal strength function}

To apply a statistical approach to compute cross sections and rates for neutrino reactions in hot nuclei, we introduce the thermal strength function for the transition operator~$\mathcal{T}$:
\begin{align}
  S_\mathcal{T}(E,T)=&\sum_{i,f}S_{if}(\mathcal{T})\delta(E-E_f+E_i)\frac{z_i}{Z},
\end{align}
where $z_i=\exp(-E_i/T)$, $Z(T)=\sum_i z_i$ is the partition function, $S_{if}(\mathcal{T})=|\langle f|\mathcal{T}|i\rangle|^2$ is the transition strength, and $E_i$, $E_f$ denote the initial and final nuclear energies. The strength function $S_\mathcal{T}(E,T)$ is determined for both positive and negative transition energies and it obeys the principle of detailed balance,
\begin{equation}\label{balance}
  S_\mathcal{T}(-E,T)=S_\mathcal{T}(E,T)\exp\Bigl(-\frac{E}{T}\Bigr).
\end{equation}
This relation links the probabilities to transfer and gain energy from a hot nucleus.

To compute the thermal strength function within the TFD formalism,  we need to know the eigenstates and eigenvalues of the so-called thermal Hamiltonian $\mathcal{H}$. The latter is defined as the difference between the physical nuclear Hamiltonian $H(a^\dag,a)$ and an auxiliary Hamiltonian $\widetilde H\equiv H(\widetilde a^\dag,\widetilde a)$ which is acting in an independent Hilbert space,
\begin{equation}
  \mathcal{H}=H-\widetilde H.
\end{equation}
By construction, the thermal Hamiltonian has both positive- and negative-energy eigenstates, $\mathcal{H}|n\rangle=E_n|n\rangle$ and $\mathcal{H}|\widetilde n\rangle=-E_n|\widetilde n\rangle$.
The zero-energy eigenstate $|0(T)\rangle$ of the thermal Hamiltonian, which satisfies the thermal state condition
\begin{equation}\label{TSC}
A|0(T)\rangle = \sigma_A\,{\rm e}^{{\mathcal H}/2T} {\widetilde
A}^\dag|0(T)\rangle,
\end{equation}
is called the thermal vacuum. It describes the equilibrium properties of the hot system. This means that the thermal average for any operator is given by the expectation value $\langle 0(T)|A|0(T)\rangle$. In~Eq.~\eqref{TSC},  $\widetilde A$ is a tilde partner of the physical operator $A$, and $\sigma_A$ is a phase factor. The correspondence between physical operators and their tilde-partners is given by the tilde-conjugation rules~\cite{Takahashi_IJMPB10,Umezawa1982,Ojima_AnPhys137}.

Given the eigenstates of the thermal Hamiltonian, the thermal strength function can be written as
\begin{equation}\label{str_func_TFD}
  S_\mathcal{T}(E,T)=\sum_n\bigl\{ S_n(\mathcal{T})\delta(E_n-E)+\widetilde S_n(\mathcal{T})\delta(E_n+E)\bigr\},
\end{equation}
where $S_n(\mathcal{T})$ and $\widetilde S_n(\mathcal{T})$ are the transition strengths
\begin{align}
  S_n(\mathcal{T}) = |\langle n|\mathcal{T}|0(T)\rangle|^2,
  \notag\\
  \widetilde S_n(\mathcal{T}) = |\langle\widetilde n|\mathcal{T}|0(T)\rangle|^2.
\end{align}
The transition strengths from the thermal vacuum to tilde-conjugated eigenstates (i.e., $|n\rangle$ and $|\widetilde n\rangle$)  of the thermal Hamiltonian  are connected by the relation
\begin{equation}
 \widetilde S_n(\mathcal{T}) =  S_n(\mathcal{T})\exp\Bigl(-\frac{E_n}{T}\Bigr),
\end{equation}
which yields the principle of detailed balance~\eqref{balance}.

Obviously, in most practical cases one cannot diagonalize $\mathcal{H}$ exactly. In the present study we apply the Thermal Quasiparticle Random-Phase-Approximation (TQRPA) and diagonalize the thermal Hamiltonian in terms of thermal phonon operators. Below we briefly outline the method, while the details can be found in~\cite{Dzhioev_IJMPE18} and in the Appendix.

\subsection{TQRPA with finite-rank separable approximation for the Skyrme interaction}

Applying the TQRPA to obtain the thermal GT$_0$ strength function we suppose that the nuclear proton and neutron Hartree-Fock (HF) states are already produced when using the Skyrme energy density functional. In particular, it means that we ignore the influence of temperature on the nuclear mean field. Following Ref.~\cite{Bortignon_1998}, this stability of the mean field with respect to temperature is expected for $T$ considerably smaller than the energy difference between major shells ($\hbar\omega_0 = 41A^{-1/3}$). Thus, the model Hamiltonian has the form
\begin{equation}
  H = H_\mathrm{mf}+H_\mathrm{pair}+H_\mathrm{ph}
\end{equation}
and it contains a spherical Skyrme-HF mean field for nucleons, the pairing interaction and the residual particle-hole interaction defined in terms of second derivatives of the Skyrme energy density functional with respect to the one-body density~\cite{Bertsch_PRepC18}. The particle-hole interaction can be written in the language of
Landau-Migdal theory of Fermi systems. Keeping only $l=0$ terms in  $H_\mathrm{ph}$, the spin part of the residual interaction reads
\begin{equation}\label{S_Landau}
  H^S_\mathrm{ph}=N_0^{-1}[G_0\sigma_1\cdot\sigma_2 + G_0^\prime\sigma_1\cdot\sigma_2\tau_1\cdot\tau_2]\delta(\mathbf{r}_1-\mathbf{r}_2),
\end{equation}
where $\sigma$ and $\tau$ are the nucleon spin and isospin operators, and $N_0=2k_Fm^*/\pi^2\hbar^2$ with $k_F$ and $m^*$ denoting the Fermi momentum and nucleon effective mass, respectively. The expressions for the Landau parameters $G_0$, $G^\prime_0$ in terms of the Skyrme force parameters can be found in Ref.~\cite{Giai_PLB379}. Because of the density dependence of the Skyrme interaction, the Landau parameters are functions of the coordinate~$\mathbf{r}$.

Following the method presented in Refs.~\cite{Giai_PRevC57,Sever_PRC66},  we apply an $N$-point integration Gauss formula and reduce $H^S_\mathrm{ph}$  to a finite-rank separable form
\begin{align}\label{V}
  H^S_\mathrm{ph} = -\frac12\sum_{k=1}^{N}\sum_{LJM}
  \sum_{\genfrac{}{}{0pt}{1}{\tau=n,p}{\rho=\pm1}}
   (\kappa^{(k)}_0\!+\!\rho\kappa^{(k)}_1) S^{(k)\dag}_{LJM}(\tau)S^{(k)}_{LJM}(\rho\tau).
\end{align}
Here, the summation is performed over the proton ($\tau=p$) and neutron ($\tau=n$) indices, and changing  the sign of isotopic index $\tau\leftrightarrow-\tau$ means changing $p\leftrightarrow n$. The isoscalar and isovector interaction strengths, $\kappa^{(k)}_0$ and $\kappa^{(k)}_1$, are expressed via the Landau parameters~~\cite{Giai_PRevC57,Sever_PRC66}. The spin-multipole operators entering $H^S_\mathrm{ph}$ are given by~\footnote{In~\eqref{SM} and hereinafter, $[]^J_M$ denotes the coupling of two single-particle angular momenta $j_1$, $j_2$ to the angular momentum $J$. The bar over index $j$ implies time inversion.}
\begin{align}\label{SM}
  S^{(k)\dag}_{LJM}(\tau) &= \hat
J^{-1}{\sum_{j_1j_2}}^\tau g^{(LJk)}_{j_1j_2}[a^\dag_{j_1}a_{\overline{\jmath_2}}]^J_M,
\end{align}
where $\hat J=\sqrt{2J+1}$  and $g^{(LJk)}_{j_1j_2}$ denotes the reduced single-particle matrix element
\begin{equation}
g^{(LJk)}_{j_1j_2}=u_{j_1}(r_k)u_{j_2}(r_k)i^L\langle j_1\|[Y_L\times\sigma]^M_J\|j_2\rangle.
\end{equation}
The radial wave functions $u_j(r_k)$ are related to the Hartree-Fock single-particle wave functions~\cite{Giai_PRevC57,Sever_PRC66}, while $r_k$ are abscissas used in the $N$-point integration Gauss formula. In Eq.~\eqref{SM}, $\sum^\tau$ implies a summation over neutron or proton single-particle states only.

Following the TFD prescription for the response functions of a hot nucleus we have to double the original nuclear degrees of freedom by introducing creation and annihilation operators acting in the auxiliary (tilde) Hilbert space and then diagonalize the respective thermal Hamiltonian. Within the TQRPA, the thermal Hamiltonian is diagonalized in two steps. First, we introduce thermal quasiparticles that diagonalize the mean field and pairing parts of $\mathcal{H}$
\begin{equation}
  \mathcal{H}_\mathrm{mf}+\mathcal{H}_\mathrm{pair} \simeq \sum_{\tau=n,p}{\sum_{jm}}^\tau \varepsilon_{jm}(T)(\beta^\dag_{jm}\beta_{jm}-
  \widetilde{\beta}^\dag_{jm}\widetilde{\beta}_{jm}).
\end{equation}
Then, to account for the residual particle-interaction we diagonalize the thermal Hamiltonian in terms of thermal phonon creation and annihilation operators:
\begin{equation}
{\cal H}\simeq\sum_{JM i}\omega_{J i}(T)
   (Q^\dag_{JM i}Q^{\phantom{\dag}}_{JM i}
   -\widetilde Q^\dag_{JM i}\widetilde Q^{\phantom{\dag}}_{JM i}).
\end{equation}
Their vacuum is the thermal vacuum in the TQRPA approximation. The energy and the structure of thermal phonons are obtained by the solution of TQRPA equations. An explicit form of the TQRPA equations for the finite-rank separable Skyrme forces is given in the Appendix, where we also discuss thermal effects on the structure and energy of the thermal phonons. Here we just note that the phonon spectrum at finite temperature contains low- and negative-energy states  which describe transitions between nuclear excited states. It should be stressed  that, in the zero-temperature limit, the TQRPA method reduces into the standard QRPA.

Once the energy and the structure of $J^\pi = 1^+$ thermal phonons are determined, one can evaluate the GT$_0$ transition strengths from the thermal vacuum to one-phonon states
 \begin{align}\label{trans_ampl}
S_i(\mathrm{GT}_0)&=\bigl|\langle Q_{1^+i}\|\sigma t_0\|0(T)\rangle\bigr|^2,
  \notag\\
\widetilde S_i(\mathrm{GT}_0)&=\bigl|\langle \widetilde Q_{1^+i}\|\sigma t_0\|0(T)\rangle\bigr|^2.
\end{align}
Substituting, $S_i$ and $\widetilde S_i$ into Eq.~\eqref{str_func_TFD} we get the thermal GT$_0$ strength function within the TQRPA.

\section{Results}\label{results}

In this section we employ the theoretical framework described above to study thermal effects on the neutral-current neutrino reactions for the two sample nuclei, $^{56}\mathrm{Fe}$ and $^{82}\mathrm{Ge}$. The iron isotope is among the most abundant nuclei at the early stages of the core-collapse, while the neutron-rich germanium isotope can be considered as the average nucleus at later stages~\cite{Cooperstein_NPA420}.

In order to estimate the  sensitivity of our results with respect to the choice of Skyrme forces, the calculations are performed by three various Skyrme parametrizations, SLy4, SGII and SkM*.
SLy4~\cite{Chabanat_NPA635} is one of the most successful Skyrme forces and has been extensively used in recent years. The  force SGII~\cite{Giai_PLB379} has been successfully applied to study spin-isospin excitations in spherical and deformed nuclei. The SkM* force~\cite{Bartel_NPA79} is an example of the first-generation Skyrme parametrizations. In what follows we  compare the results of TQRPA calculations with the Skyrme functionals with those performed within the QPM Hamiltonian~\cite{Dzhioev_PRC89,Dzhioev_PAN77}. To distinguish between the two approaches, we  refer to them as QPM-TQRPA and Skyrme-TQRPA.

Let us also make a short remark concerning the choice of the pairing interaction. Within the BCS approach the phase transition in nuclei from the superfluid to normal state occurs at critical temperatures ($T_\mathrm{cr}\approx0.5\Delta$, where $\Delta$ is a pairing gap)~\cite{Goodman_NPA352}. Therefore, the inclusion of particle-particle residual interactions does not affect the strength function for temperatures $T> T_\mathrm{cr}$. However, to compare our results with the  shell-model calculations, pairing correlations are taken into account at zero temperature. As in Refs.~\cite{Dzhioev_PRC89,Dzhioev_PAN77}  we employ in the present study a BCS Hamiltonian with a constant pairing strength. The neutron and proton pairing strength parameters are fixed to reproduce the odd-even mass difference. At $T=0$ the resulting proton and neutron energy gaps are $\Delta_{p(n)}=1.57(1.36)$~MeV for $^{56}$Fe and $\Delta_{p(n)}=1.22(0)$~MeV for $^{82}$Ge. Thus, the critical temperature  when the pairing phase transition occurs in the iron isotope is $T_\mathrm{cr}\approx 0.8$~MeV and $T_\mathrm{cr}\approx 0.6$~MeV for the germanium isotope.

\subsection{GT$_0$ strength functions at zero and finite temperature}

\begin{figure}[t]
 \begin{centering}
\includegraphics[width=1.0\columnwidth]{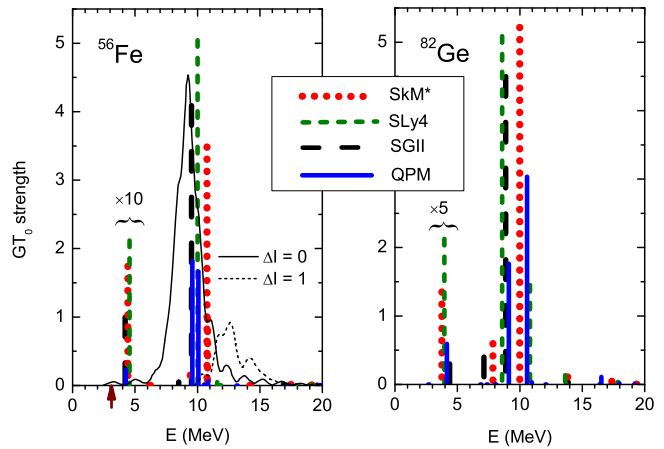}
\caption{(Color online) The ground-state GT$_0$ strength distributions for $^{56}$Fe and $^{82}$Ge. The Skyrme-QRPA calculations are performed with the SkM*, SLy4, and SGII energy density functionals. The GT$_0$ strength calculated in Ref.~\cite{Dzhioev_PRC89} with the QPM Hamiltonian are also shown. Note that, for a clearer presentation, the low-energy peaks in $^{56}$Fe and $^{82}$Ge  are scaled, i.e., the values displayed are obtained by multiplying the actual values by the indicated factor. The shell-model GT$_0$ strength (scaled by a factor of 3.5) from Ref.~\cite{Sampaio_PLetB529} is shown for $^{56}$Fe. The shell-model strength is split into the two isospin components: $\Delta I = 0$ (full line) and $\Delta I = 1$ (dashed line). The arrow indicates the experimental energy of the first $1^+$ state in $^{56}$Fe ($E_\mathrm{exp}(1^+_1)=3.12$~MeV). This energy corresponds to
the threshold for inelastic neutrino scattering from the ground state of $^{56}$Fe.}
 \label{GT0_gs}
 \end{centering}
\end{figure}

Before considering neutral-current neutrino reaction with hot nuclei, let us first analyze the thermal effects on the GT$_0$ strength functions in $^{56}\mathrm{Fe}$ and $^{82}\mathrm{Ge}$. In Ref.~\cite{Dzhioev_PRC89} such an analysis was performed by applying the QPM-TQRPA. Here we repeat the same calculations by using the Skyrme-TQRPA and compare the results of two approaches.

\begin{table}[t!]
    \caption{Total GT$_0$ strength, $S(\mathrm{GT}_0$), calculated with different residual forces. In parentheses, the unperturbed ($H_\mathrm{ph}=0$) values are shown.  }
  \begin{tabular}{ccccc}
    \hline\hline
      & SGII~~ & SkM*~~ & SLy4~~ & QPM \\
    \hline\hline
    $^{56}$Fe~~~  & 4.3 (5.1)~~~  & 4.9 (5.5)~~~ & 5.4 (5.5)~~~ & 3.6 (5.0)   \\
    \hline
    $^{82}$Ge~~~  & 5.1 (5.9)~~~  & 6.3 (7.1)~~~ & 7.2 (7.3)~~~ & 5.2 (7.0)   \\
     \hline\hline
  \end{tabular}
 \label{total_strength}
\end{table}

In Fig.~\ref{GT0_gs} we display the ground-state GT$_0$ strength distributions for $^{56}\mathrm{Fe}$ and $^{82}\mathrm{Ge}$ obtained within the QRPA with the various Skyrme forces. For comparison, the strength distributions obtained in Ref.~\cite{Dzhioev_PRC89} with the QPM Hamiltonian are also plotted. Here, we would like to remind the reader, that in Ref.~\cite{Dzhioev_PRC89}, the strength parameters of the residual interaction in $^{56}\mathrm{Fe}$ were fit to reproduce the experimental energy centroids of the GT$_-$ and GT$_+$ resonances~\cite{Rapaport_NPA410,Ronnqvist_NPA563}, while for $^{82}\mathrm{Ge}$ the strength parameters were estimated following the parametrizations given in Refs.~\cite{Castel_PLetB65,Bes_PRep16}.

As seen in Fig.~\ref{GT0_gs} all the three Skyrme forces produce very similar GT$_0$ strength distributions. Namely,  a major part of the strength concentrates in a resonance peak  around 10~MeV and there is some smaller strength at low energies. According to the present Skyrme-QRPA calculations, the resonance peak in $^{56}$Fe is a superposition of the neutron and proton spin-flip transitions $1f_{7/2}\to 1f_{5/2}$, while  the low-energy peak is formed by the $2p_{3/2}\to 2p_{1/2}$ neutron transition. In $^{82}$Ge, the neutrons fully occupy the $2p_{3/2,\,1/2}$, $1f_{7/2,5/2}$ and $1g_{9/2}$ orbits. Therefore, the principal contribution to the GT$_0$ resonance comes from the proton $1f_{7/2}\to 1f_{5/2}$ and the neutron $1g_{9/2}\to 1g_{7/2}$ transitions, whereas the low-energy strength results from the  proton $2p_{3/2}\to 2p_{1/2}$ transition.

In Table~\ref{total_strength} we give the summed GT$_0$ strengths,  $S(\mathrm{GT}_0) =\sum S_i(\mathrm{GT}_0$), calculated by using the different Skyrme functionals with and without taking into account the residual interaction. As can be seen from the table, the particle-hole correlations reduce the total GT$_0$ strength. This collective effect is most significant for the SGII Skyrme functional. Note also that SGII predicts the smallest low-energy GT$_0$ peak as compared to SkM* and SLy4.

Referring to Fig.~\ref{GT0_gs}, the results of QPM-QRPA calculations are in a good agreement with the Skyrme-QRPA calculations in the sense that the most part of the GT$_0$ strength is concentrated around 10~MeV and there is a small peak around $4$~MeV. We note, however, that in both nuclei the phenomenological Woods-Saxon mean field and residual interaction yield a two-peak structure of the GT$_0$ resonance, while the self-consistent calculations with Skyrme forces predict a one-peak structure. Moreover, as seen from Table~\ref{total_strength}, the QPM-QRPA gives a significant (about 25\%) reduction of the total GT$_0$ strength due to particle-hole correlations. In  $^{56}$Fe this reduction leads to a somewhat lower value of  $S(\mathrm{GT}_0)$ as compared with the Skyrme-QRPA values. Furthermore, in $^{56}$Fe the particle-hole correlations induced by the phenomenological residual interaction strongly suppress the low-energy peak of the GT$_0$ distribution.

In the left panel of Fig.~\ref{GT0_gs}, the QRPA distribution is compared with the LSSM calculations for~$^{56}$Fe~\cite{Sampaio_PLetB529}. One notices that the gross structure of the QRPA and LSSM distributions agrees well with each other. In particular, SGII very accurately reproduces the shell-model position of the GT$_0$ resonance. However, since the QRPA fails to recover all nuclear correlations needed to correctly describe the fragmentation of the strength, the fine structure of the GT$_0$ distributions in the vicinity of the resonance is not reproduced in our calculations. Note also that the QRPA calculations systematically predict the energy of the lowest excited state in $^{56}$Fe to be higher than the experimental value. In this respect the shell-model calculations are clearly advantageous.

\begin{figure}[t]
 \begin{centering}
\includegraphics[width=1.0\columnwidth]{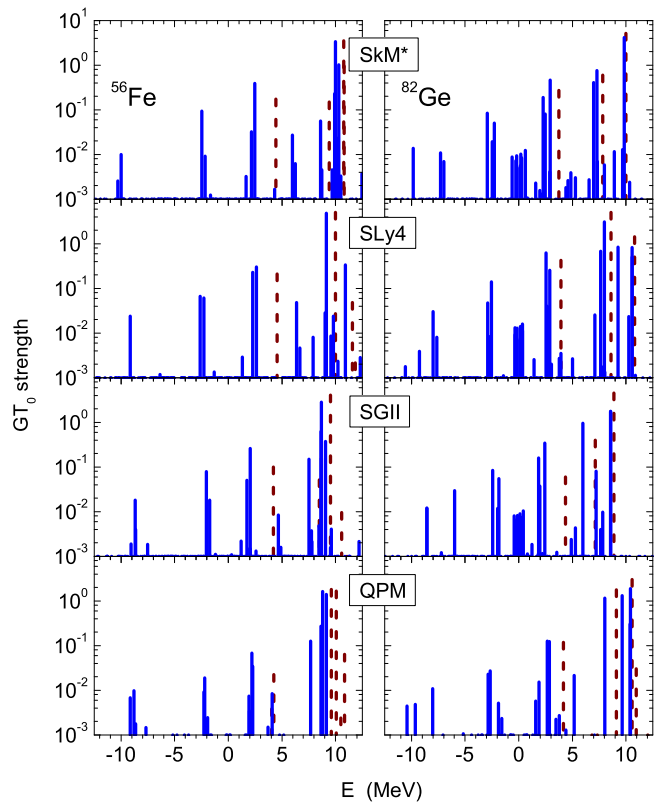}
\caption{(Color online) The GT$_0$ strength function for $^{56}$Fe and $^{82}$Ge calculated at $T=0$ (dashed peaks) and $T=1.72$~MeV  (solid peaks).     }
 \label{GT0_Temp}
 \end{centering}
\end{figure}

Let us now compare the results of Refs.~\cite{Dzhioev_PRC89,Dzhioev_PAN74} where thermal effects on the GT$_0$ strength functions were studied within the QPM-TQRPA approach with the present self-consistent scheme based on the Skyrme interaction, i.e., with the Skyrme-TQRPA. In Fig.~\ref{GT0_Temp}, the ground-state GT$_0$ strength functions in $^{56}$Fe and $^{82}$Ge are compared with those obtained at a finite temperature of $T=1.72$~MeV. This temperature roughly corresponds to the neutrino thermalization stage of the core collapse. Note that, to make the thermal effects more pronounced, the strength functions are displayed on a logarithmic scale.

\begin{figure}[t]
 \begin{centering}
\includegraphics[width=1.0\columnwidth]{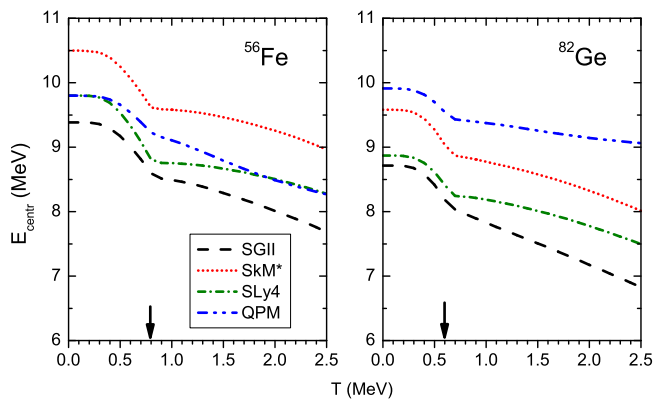}
\caption{(Color online) The energy-centroid for the upward GT$_0$ strength functions in $^{56}$Fe and $^{82}$Ge as a function of temperature. The arrows indicate the critical temperature, $T_\mathrm{cr}$, when the pairing collapses.}
 \label{GT0_centr}
 \end{centering}
\end{figure}

Since the TQRPA calculations do not employ the Brink hypothesis, we observe an evolution of the GT$_0$ strength function with temperature and this result is valid regardless of  the effective interaction. For the upward ($E>0$) transitions the main effect is a shift of the GT$_0$ strength towards lower energies. This shift is due to two reasons: First, the resonance energy is lowered because of the vanishing of pairing correlations and a thermal weakening of the residual interaction. The latter occurs due to the thermal occupation factors that appear in the TQRPA matrix elements [see Eq.~\eqref{MatEl} in the Appendix]. Since there is no neutron pairing correlation in $^{82}$Ge, the resonance lowering is more pronounced in the iron isotope. Second, the thermal smearing of the  Fermi surfaces makes possible low-energy particle-particle and hole-hole transitions which are Pauli blocked at $T=0$. Such thermally unblocked transitions enhance the low-energy component of the GT$_0$ strength and make it more fragmented.

To examine qualitatively the thermal effects on the upward GT$_0$ strength, we introduce the energy-centroid,  $E_\mathrm{centr}$, as the ratio of the first and the zeroth moments of the positive-energy strength function.
In Fig.~\ref{GT0_centr},  the  energy-centroids are shown as  functions of temperature. One can see that all Skyme forces give generally similar results for the temperature dependence of $E_\mathrm{centr}$. Namely, up to $T_\mathrm{cr}$ the temperature dependence of the energy-centroid resembles that of the pairing gap, which at low temperatures has a temperature-independent plateau and then decreases with $T$~\cite{Goodman_NPA352}.
After $T_\mathrm{cr}$, the decrease of $E_\mathrm{centr}$  becomes gradual and results from the weakening of the residual interaction and the thermal unblocking of low-energy GT$_0$ transitions. Depending on the effective interactions, when the temperature rises to $T=2.5$~MeV, the total downward shift of $E_\mathrm{centr}$ varies from 1.5 to 1.8~MeV in $^{56}$Fe and from 1.0 to 1.8~MeV in $^{82}$Ge. Note also, that among the Skyrme forces we use, the SkM* parametrization predicts the highest value for $E_\mathrm{centr}$, while the SGII force predicts the lowest value for $E_\mathrm{centr}$.

\begin{figure}[t]
 \begin{centering}
\includegraphics[width=1.0\columnwidth]{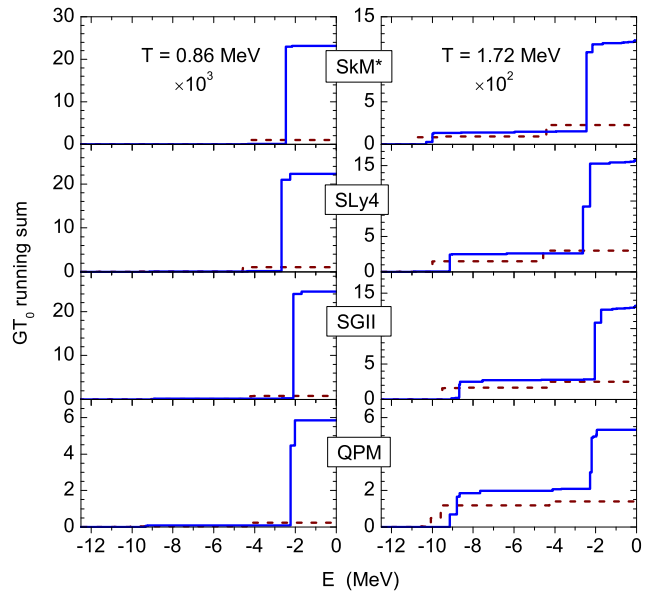}
\caption{(Color online) Comparison of the running sums for the GT$_0$ downward strength in $^{56}$Fe obtained by using  and without using the Brink hypothesis (dashed and solid lines, respectively). Note that the values are scaled by a factor of $10^3$ ($T=0.86$~MeV) and $10^2$ ($T=1.72$~MeV). }
 \label{GT0_Temp_running}
 \end{centering}
\end{figure}

As is evident from Fig.~\ref{GT0_Temp}, finite temperature also affects the strength function for downward ($E<0$) GT$_0$ transitions. In accordance with detailed balance~\eqref{balance}, the temperature rise exponentially increases the strength of negative-energy transitions. This relates to thermally populated nuclear states which decay to states at lower energies. In particular, the strength observed at $E\approx -9$~MeV is attributed to the deexcitation of the GT$_0$ resonance. It is also clear that thermal effects on the upward GT$_0$ strength should affect the downward strength as well. Following Ref.~\cite{Dzhioev_PRC89}, we study this effect by calculating the running sums for the GT$_0$ downward strength either by using the Brink hypothesis or not. The former are obtained from the QRPA ground-state  strength function by a multiplication with the Boltzmann factor $\exp(-E/T)$. Since, our general findings for $^{56}$Fe and $^{82}$Ge are similar, we consider the results for $^{56}$Fe.
The running sums are shown in Fig.~\ref{GT0_Temp_running} for  $T=0.86$~MeV and $T=1.72$~MeV.
Within both the QPM-TQRPA and the Skyrme-TQRPA models, ignoring the Brink hypothesis leads to a considerable enhancement of the total downward strength. Moreover, regardless of the effective forces this enhancement is mainly caused by the thermal effects on the low-energy strength function. This is most pronounced at low-temperatures ($T=0.86$~MeV). However, at higher temperatures ($T=1.72$~MeV) the lowering of GT$_0$ resonance energy also contributes to the downward strength enhancement.

Referring to the figure, the total downward strengths obtained with the Skyrme-TQRPA considerably exceed those obtained with the QPM-TQRPA calculations:  at $T=0.86$~MeV the difference is about a factor of 4, while  at $T=1.72$~MeV the difference is about a factor of 3. As evident from Figs.~\ref{GT0_Temp} and~\ref{GT0_Temp_running}, this discrepancy comes from the facts that the Skyrme-TQRPA models predict larger strengths for low-energy GT$_0$ transitions.

\subsection{Inelastic neutrino scattering}

\begin{figure}[t]
 \begin{centering}
\includegraphics[width=1.0\columnwidth]{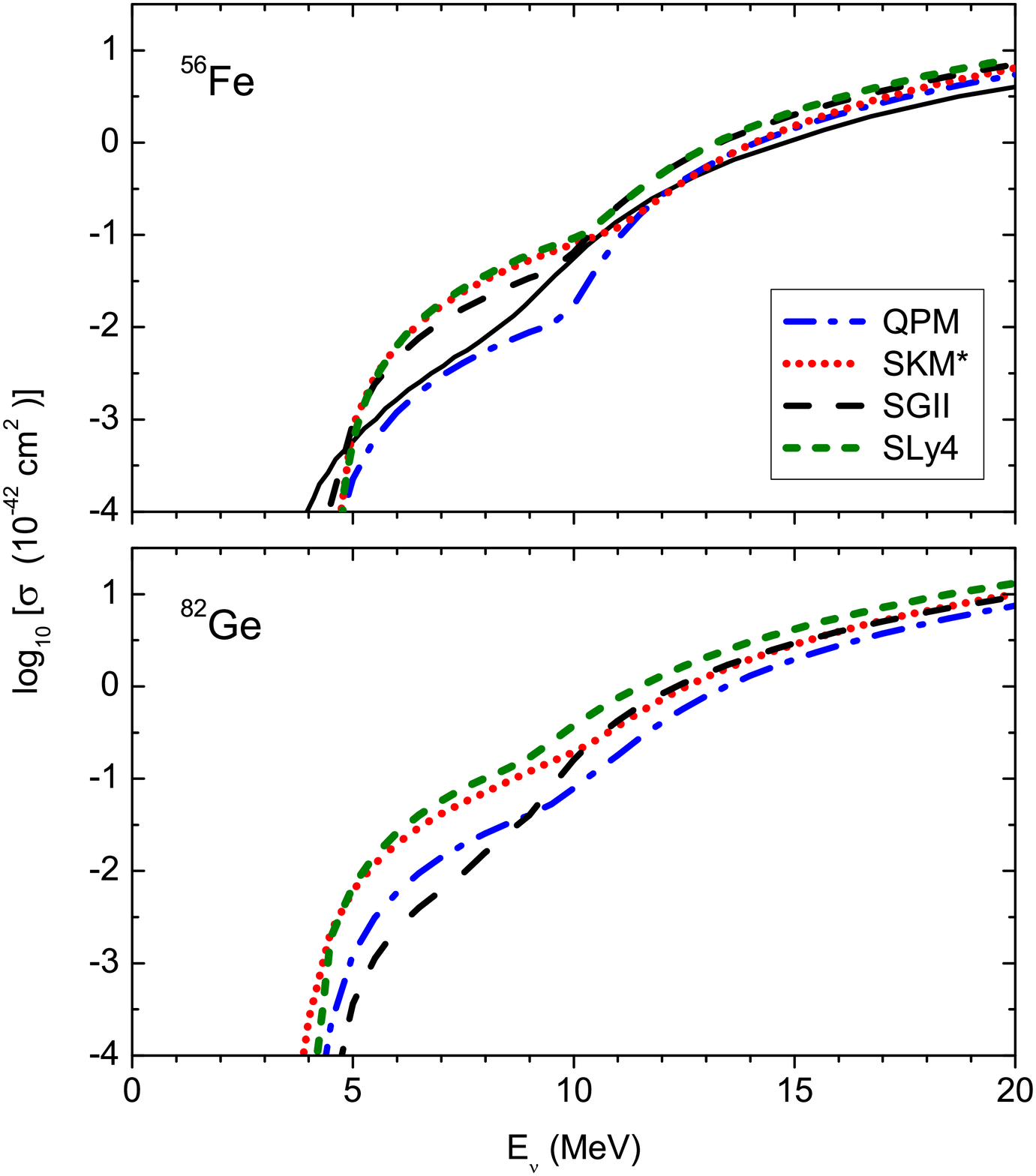}
\caption{(Color online) Inelastic neutral-current neutrino scattering cross sections from the ground states of $^{56}$Fe and $^{82}$Ge. The cross sections are calculated within the Skyrme-QRPA (SkM*, SGII, Sly4) as well as the QPM-QRPA~\cite{Dzhioev_PRC89}. For $^{56}$Fe, the results of shell-model calculations~\cite{Sampaio_PLetB529} are shown by the solid line.}
 \label{CrSect_gs}
 \end{centering}
\end{figure}

In this section we calculate inelastic neutrino scattering cross-sections for the sample nuclei and compare the results with those obtained within the QPM-TQRPA model~\cite{Dzhioev_PRC89} and those
of Ref.~\cite{Juodagalvis_NPA747}, where the hybrid model  has been employed. As was mentioned in the introduction, the hybrid model takes into account the GT$_0$ distributions from the LSSM, while other multipole contributions
are derived from the RPA calculation. At energies $E_\nu < 20$~MeV forbidden contributions are at least an order of magnitude smaller than the GT contributions.

Considering only GT$_0$ transitions, the cross section for neutrino scattering from a hot nucleus reads
\begin{align}\label{CrSect}
  \sigma(E_\nu, T)&=\frac{G^2_F }{\pi}(g^\mathrm{eff}_A)^2\sum_{i,f} (E_{\nu'})^2 |\langle f\|\sigma t_0\|i\rangle|^2\frac{z_i}{Z}
   \notag\\
  &=\sigma_\mathrm{down}(E_\nu, T)+\sigma_\mathrm{up}(E_\nu, T).
\end{align}
Here, $G_F$ is the Fermi coupling constant, $g^\mathrm{eff}_A$ is the effective axial coupling constant which simulates the observed quenching of the GT strength at energies $E_x \lesssim 15-20$~MeV,  and $E_{\nu'} = E_\nu - (E_f-E_i)$ is the energy of outgoing neutrino. The down-scattering component of the cross section, $\sigma_\mathrm{down}(E_\nu, T)$, includes only upward transitions ($E_f>E_i$), while downward transitions ($E_i>E_f$) are included into the up-scattering term $\sigma_\mathrm{up}(E_\nu, T)$. We recall that the hybrid approach of Ref.~\cite{Juodagalvis_NPA747} explicitly incorporates  Brink's hypothesis which assumes that upward GT$_0$ transitions from excited states have the same strength distribution as those from the nuclear ground state. Under this assumption the down-scattering cross section becomes temperature independent.  However, this is likely not the case
 because the vanishing of pairing correlations and thermal smearing of the Fermi surface should affect the distribution centroid and move it down in energy. In the above discussion, this effect was clearly demonstrated within  the TQRPA. In Ref.~\cite{Juodagalvis_NPA747}, the finite-temperature effects are included only into
the up-scattering cross section by inverting the shell-model GT$_0$ distributions for the lowest excited states .

By expressing the cross section~\eqref{CrSect} through the thermal strength function
\begin{align}
\sigma(E_\nu, T) =\frac{G^2_F}{\pi}(g^\mathrm{eff}_A)^2\int (E -E_\nu)^2 S_\mathrm{GT}(E,T)dE
\end{align}
and then applying the TQRPA we get
\begin{align}
  \sigma(E_\nu, T)=&\frac{G^2_F }{\pi}(g^\mathrm{eff}_A)^2\biggl\{{\sum_i}'(E_\nu - \omega_i)^2 S_i
  \notag\\
  &+\sum_i(E_\nu + \omega_i)^2 S_i\exp\Bigl(-\frac{\omega_i}{T}\Bigr)\biggr\}.
\end{align}
Here, the first term corresponds to the down-scattering contribution and it implies summation over $1^+$ thermal-phonon states with positive energy $\omega_i<E_\nu$. The second, up-scattering term, accounts for GT$_0$ transitions to negative-energy states. In contrast to the hybrid model, in the TQRPA both the down- and up-scattering terms are temperature dependent.

In Fig.~\ref{CrSect_gs} we display the ground-state cross sections as functions of the incident neutrino energy $E_\nu$.  For both nuclei the Skyrme-QRPA results are compared with the cross sections calculated by using the QPM Hamiltonian~\cite{Dzhioev_PRC89}. For $^{56}$Fe  the comparison is also made with the shell-model cross section~\cite{Sampaio_PLetB529}. To make the comparison with the shell-model calculations more transparent, we use the same reduction of the axial-weak coupling constant from its free-nucleon value $g_A=-1.2599$~\cite{Towner1995} to the effective value  $g_A^\mathrm{eff}=0.74g_A$.

Referring to the figure, the cross sections exhibit a sharp increase within the first few MeV above threshold. For neutrino energies $E_\nu>10$, where the excitation of the GT$_0$ resonance is possible, the increase becomes more gradual. Even if the general trend of the Skyrme-QRPA cross sections as a function of the neutrino energy is in agreement with the QPM and the shell-model results, the absolute values differ by up to an order of magnitude. The disagreement is most pronounced at low neutrino energies, $E_\nu\lesssim 10$~MeV, where the cross sections are very sensitive to the details of the GT$_0$ strength distribution. In $^{56}$Fe, at low energies ($5\lesssim E_\nu\lesssim10$~MeV) all Skyrme-QRPA cross sections are considerably above the values obtained in the QPM-QRPA and LSSM calculations. The reason for this systematic effect is that, in the iron isotope, the Skyrme-based calculations predict a larger strength of low-energy GT$_0$ transitions than the other two approaches~(see Fig.~\ref{GT0_gs}).  For the same reason the low-energy cross sections for $^{82}$Ge calculated with the SkM* and SLy4 forces are larger than those obtained with the SGII and phenomenological spin-spin forces. With increasing neutrino energy the spread in the cross sections is reduced and for $E_\nu>15$~MeV all Skyrme calculations give very similar cross sections. Note, however, that for $E_\nu>15$~MeV neutrinos the Skyrme-QRPA model predicts cross sections slightly above the values of other two models. It is clear that this discrepancy reflects the differences in the total GT$_0$ strength and in the energy centroid.

\begin{figure}[t]
 \begin{centering}
\includegraphics[width=1.0\columnwidth]{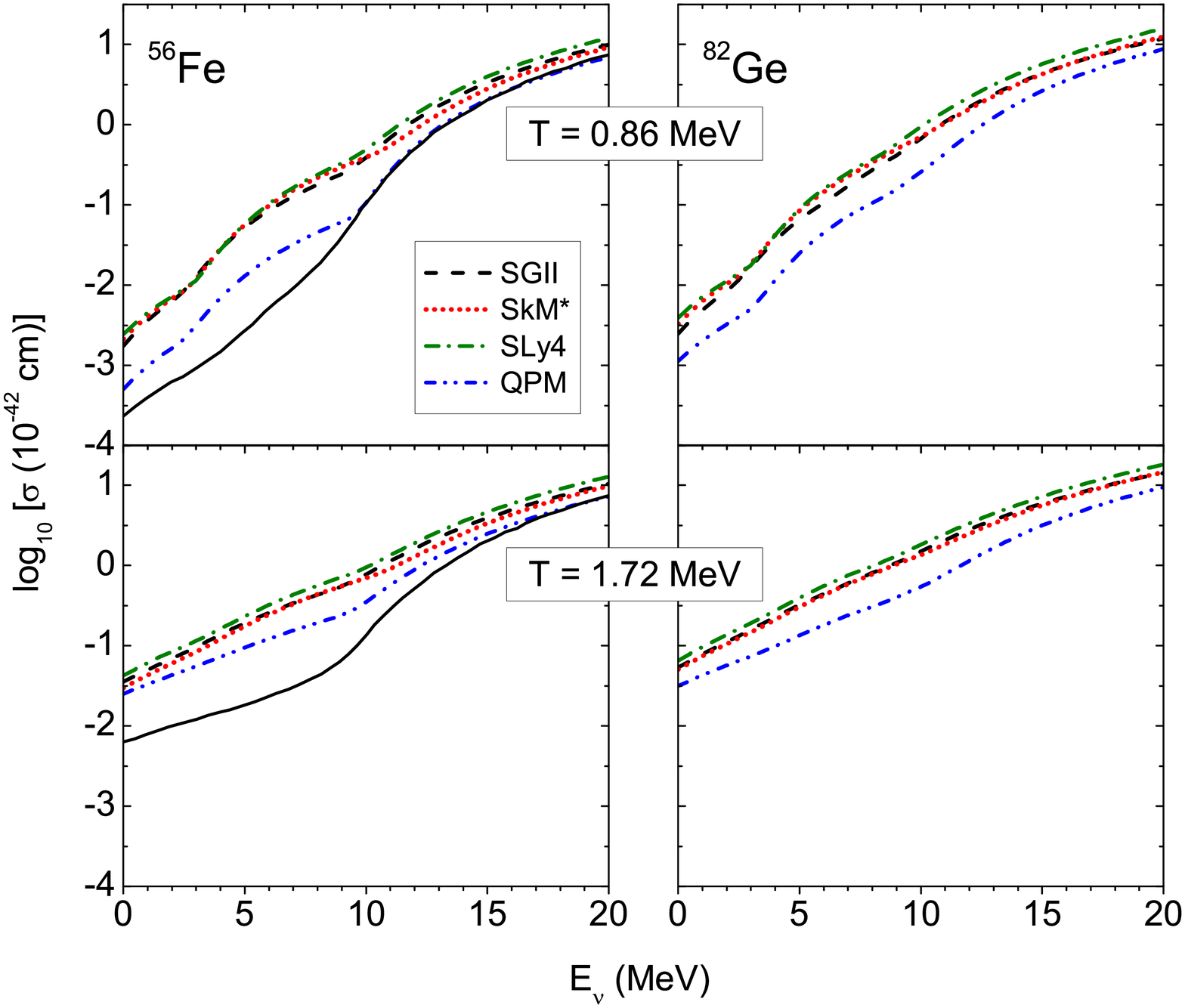}
\caption{(Color online) Inelastic neutrino scattering cross sections for $^{56}$Fe and $^{82}$Ge at $T=0.86$ and 1.72~MeV, calculated within the TQRPA. For $^{56}$Fe,
the cross sections obtained within the hybrid model~\cite{Juodagalvis_NPA747}  are shown by the solid line.}
 \label{CrSect_Temp}
 \end{centering}
\end{figure}

In Fig.~\ref{CrSect_Temp}, the neutrino-scattering cross sections are shown for $T=0.86$ and 1.72~MeV together with the results from the QPM-TQRPA model~\cite{Dzhioev_PRC89}. For $^{56}$Fe, the TQRPA results are compared with those obtained within the hybrid model~\cite{Juodagalvis_NPA747}. As seen from the plots, all models predict that there is no gap in the cross section at $T\ne 0$. This is caused by the fact that deexcitation of thermally excited states may contribute to the cross section at all neutrino energies. The contribution of thermally populated states increases with temperature and enhances the low-energy cross section. With increasing neutrino energies, thermal effects on the cross section become less important.

In Ref~\cite{Dzhioev_PRC89}, it was shown that the TQRPA  predicts a more significant thermal enhancement of the low-energy cross section as compared with  the hybrid approach.  Recall, that within the TQRPA both the down- and up-scattering contributions to the cross section [see Eq.~\eqref{CrSect}] increases with temperature, while in the hybrid model, due to employing the Brink hypothesis, only the up-scattering term is temperature-dependent and contributes to the thermal enhancement of the cross section.  As seen from Fig.~\ref{CrSect_Temp}, the TQRPA calculations with Skyrme forces allow for even a greater thermal enhancement of the cross section than those obtained with the QPM Hamiltonian. From the analysis in the previous section, we conclude that the reason for this is a larger strength of thermally unblocked low-energy GT$_0$ transitions within the Skyrme-TQRPA. It is important to note that different Skyrme functionals give very similar finite-temperature cross sections.

\begin{figure*}[t]
 \begin{centering}
\includegraphics[width=0.8\textwidth]{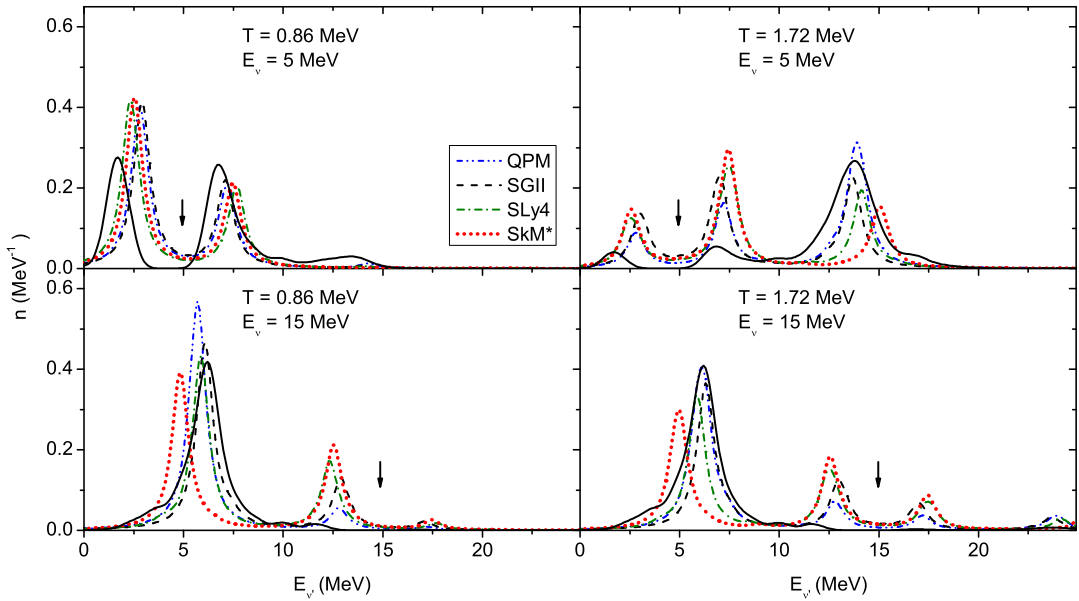}
\caption{(Color online) Normalized spectra of outgoing neutrinos for $^{56}$Fe at $T=0.86,~1.72$~MeV and two initial neutrino energies $E_\nu = 5,~15$~MeV.
  In all plots the energy of the incoming neutrino is indicated by an arrow. A comparison is made between the TQRPA results and those obtained in Ref.~\cite{Juodagalvis_NPA747} within the hybrid model. The latter are shown   by the solid lines.}
 \label{spectrum}
 \end{centering}
\end{figure*}

Keeping track of neutrino energies is a crucial ingredient of supernova simulations. For inelastic neutrino scattering the energy change depends on the incident energy and temperature. In Fig.~\ref{spectrum}, we show the normalized spectra of outgoing neutrinos scattered off $^{56}$Fe at the same temperatures as in Fig.~\ref{CrSect_Temp} and at two initial neutrino energies, $E_\nu = 5,~15$~MeV. In the figure, the energy distributions calculated within the TQRPA are compared to those of the hybrid model~\cite{Juodagalvis_NPA747}. Note, that the TQRPA spectra are folded by the Breit-Wigner function with a width of~1~MeV.

As seen from Fig.~\ref{spectrum}, for $E_\nu=5$~MeV the fraction of up-scattering neutrinos ($E_{\nu'}>E_\nu$)  gets significantly enhanced with increasing temperature due to the population of excited states. However, the spectra of up-scattered neutrinos calculated within the TQRPA and the hybrid model can differ essentially. At $T=1.72$~MeV, the hybrid model predicts that the cross section is dominated by the deexcitation of the GT$_0$ resonance, giving rise to a high-energy peak in the spectrum around $E_{\nu'}\approx14$~MeV. In the TQRPA, a temperature rise increases the strength of low-energy downward transitions. The low-energy peak in the spectrum at $E_{\nu'}\approx 7.5$~MeV corresponds
to such transitions, and it is more pronounced in the TQRPA spectra. Note, however, that even within the TQRPA calculations the relative fraction of low- and high-energy up-scattered neutrinos differ:  the ratio $n(E_{\nu'}\approx 7.5~\mathrm{MeV})/n(E_{\nu'}\approx 15~\mathrm{MeV})$ is the lowest ($\approx 0.5$) in the QPM-TQRPA, and is the highest ($\approx 1.9$) in the spectrum obtained with the SkM* functional. For the hybrid-model calculations this ratio is about 0.18.

For high-energy incoming neutrinos the up-scattering process becomes essentially irrelevant and for $E_\nu=15$~MeV the cross section is dominated by the excitation of the GT$_0$ resonance, as evidenced by low-energy peaks in the spectra around $E_{\nu'}=5$~MeV. However, the high-energy peaks in the spectra around $E_{\nu'}=12.5$~MeV show that, within the TQRPA, a notable fraction of neutrinos is down-scattered due to thermally unblocked low-energy transitions. No such transitions appear within the hybrid model because of the application of Brink's hypothesis. Like for the $E_\nu=5$~MeV case, the ratio of low-energy down-scattered neutrinos to high-energy down-scattered neutrinos varies between different TQRPA calculation. It is  highest for the QPM-TQRPA, while the SkM* functional predicts the lowest ratio.

\subsection{Neutrino pair emission}

Now we apply the formalism described in Sec.~\ref{formalism} to compute the rate for nuclear deexcitation by neutrino-antineutrino pair emission:
\begin{equation}
  (A,Z)^*\to (A,N) +\nu_k+ \overline{\nu}_k.
\end{equation}
Here,  the index $k = e,~\mu,~\tau$ corresponds to three neutrino flavors that can be produced in the decay.
Considering only  GT$_0$ transitions, the emission rate is given by~\cite{Fuller_APJ376,Fischer_PRC88}
\begin{align}
  \Lambda&= 3\lambda_0 \sum_{i,f} E^5_{\nu\overline{\nu}} |\langle i\|\sigma t_0\|f\rangle|^2\frac{z_i}{Z},
\end{align}
where $E_{\nu\overline{\nu}} = E_i-E_f$ is the energy of the emitted $\nu\overline{\nu}$-pair,
  $\lambda_0= G^2_F g^2_A/(60\pi^3\hbar^7c^6)\approx1.72\times10^{-4}\,\mathrm{s}^{-1}\mathrm{MeV}^{-5}$ , and the factor of three accounts for the three possible neutrino flavors.

Like for the neutrino-scattering case, we express the emission rate through the thermal strength function. Applying the TQRPA we obtain
\begin{align}\label{decay_rate}
  \Lambda&= 3\lambda_0\int^{\infty}_0 E^5 S_\mathrm{GT}(-E,T)dE
 \notag\\
  &=3\lambda_0\int^{\infty}_0 E^5 S_\mathrm{GT}(E,T)\exp\Bigl(-\frac{E}{T}\Bigr)dE
  \notag\\
  &=3\lambda_0\sum_i \omega_i^5 S_i\exp\Bigl(-\frac{\omega_i}{T}\Bigr)=\sum_i\Lambda_i.
\end{align}
In the emission rate, the Boltzmann factor suppresses the contribution from the high-energy tail of the strength function, while the phase factor $E^5$ suppresses the contribution from lower energies. We also introduce the ratio $\lambda_i = \Lambda_i/\Lambda$, which gives  the normalized spectra for the emitted $\nu\overline{\nu}$-pairs.

\begin{figure*}[t]
 \begin{centering}
\includegraphics[width=0.8\textwidth]{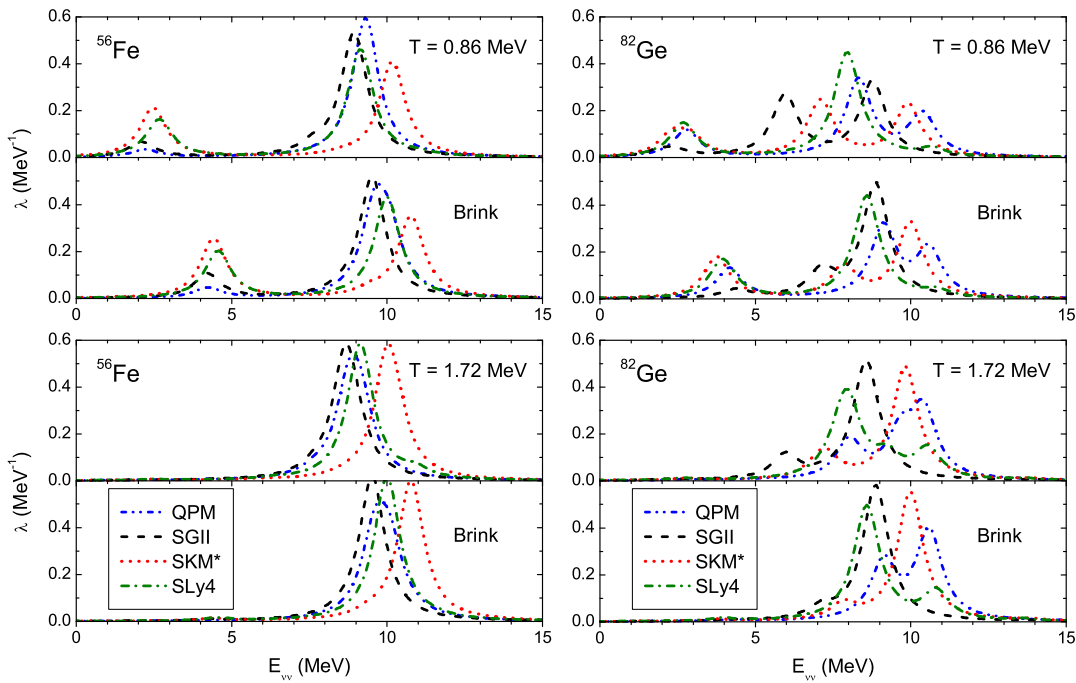}
\caption{(Color online)  Normalized spectra of $\nu\overline{\nu}$-pairs emitted by $^{56}$Fe (left panels) and $^{82}$Ge (right panels) at two different temperatures, $T=0.86$~MeV and $T=1.72$~MeV. For each temperature we show the spectra obtained  by using and without using the Brink hypothesis (lower and upper panels on each plot, respectively).}
 \label{partial_rates}
 \end{centering}
\end{figure*}

In Fig.~\ref{partial_rates} we show the normalized spectra of emitted neutrino-antineutrino pairs by  $^{56}$Fe and $^{82}$Ge at two different temperatures, $T=0.86$ and 1.72~MeV. The spectra are computed in the QPM-TQRPA and in the Skyrme-TQRPA with different Skyrme functionals. Note that for a clearer presentation the spectra are folded by the Breit-Wigner function with a width of~1~MeV. We also show the spectra obtained by applying the Brink hypothesis. In the latter case we approximate the GT$_0$ thermal strength function  in~Eq.~\eqref{decay_rate} by the QRPA ground-state GT$_0$ strength distribution.

Although the details of the spectra vary between different calculations, some essential features are the same. Namely, at $T=0.86$~MeV the spectrum is dominated by two peaks.
The peaks around $2.5$~MeV originate from thermally unblocked low-energy transitions, while the main peaks around 9~MeV  correspond to neutrino pairs emitted due to decay
of the thermally excited GT$_0$ resonance. At $T=1.72$~MeV, the low-energy pairs disappear from the spectrum  and neutrino emission is strongly dominated by the deexcitation
of the GT$_0$ resonance. It is interesting  that neutrino spectra calculated with and without application of Brink hypothesis look similar, but in the latter case they are shifted to lower energies. As seen from the figure,  the shift is more pronounced for the low-energy part of the spectrum. Note also that, in $^{82}$Ge, the spectrum obtained without applying the Brink hypothesis is somewhat broader than that obtained by using the Brink hypothesis.

\begin{figure}[t]
 \begin{centering}
\includegraphics[width=1.0\columnwidth]{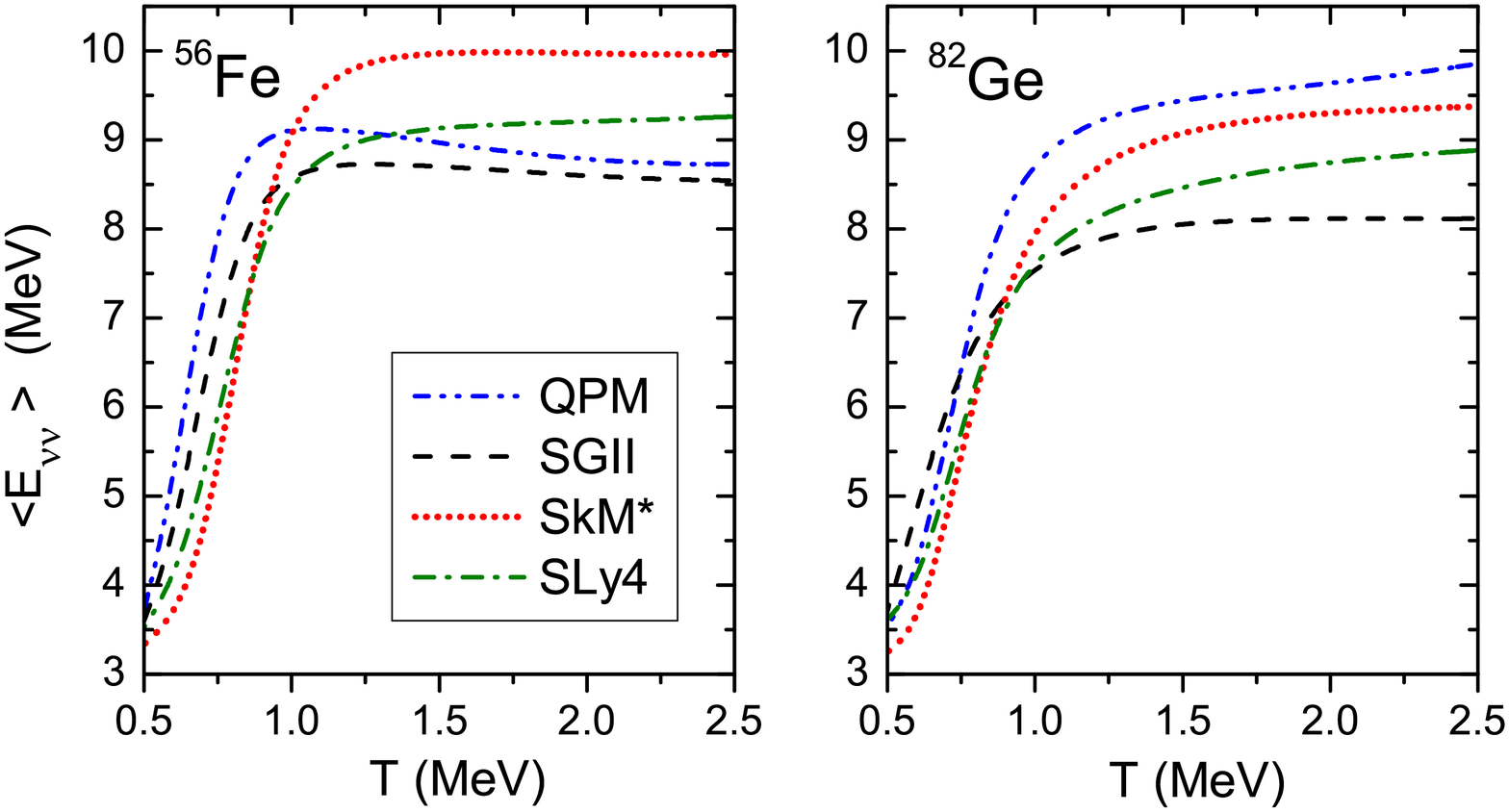}
\caption{(Color online) The energy centroid for the spectrum of emitted neutrinos as a function of temperature.}
 \label{E_mean}
 \end{centering}
\end{figure}

In Fig.~\ref{E_mean} we display the energy centroid $\langle E_{\nu\overline{\nu}}\rangle$ for the spectrum of emitted neutrino pairs as a function of temperature. As is evident from the figure, regardless of the effective forces we use, $\langle E_{\nu\overline{\nu}}\rangle$ rises rapidly with temperature untill $T\approx 1.0$~MeV. At higher temperatures,  the energy centroid becomes nearly temperature independent and its value is determined by the energy of the GT$_0$ resonance. The value of $\langle E_{\nu\overline{\nu}}\rangle$ obtained with the different Skyrme functionals varies over the energy interval between 8 and 10~MeV. For both nuclei the SkM* force gives the largest values, while SGII predicts the smallest one.

\begin{figure}[t]
 \begin{centering}
\includegraphics[width=1.0\columnwidth]{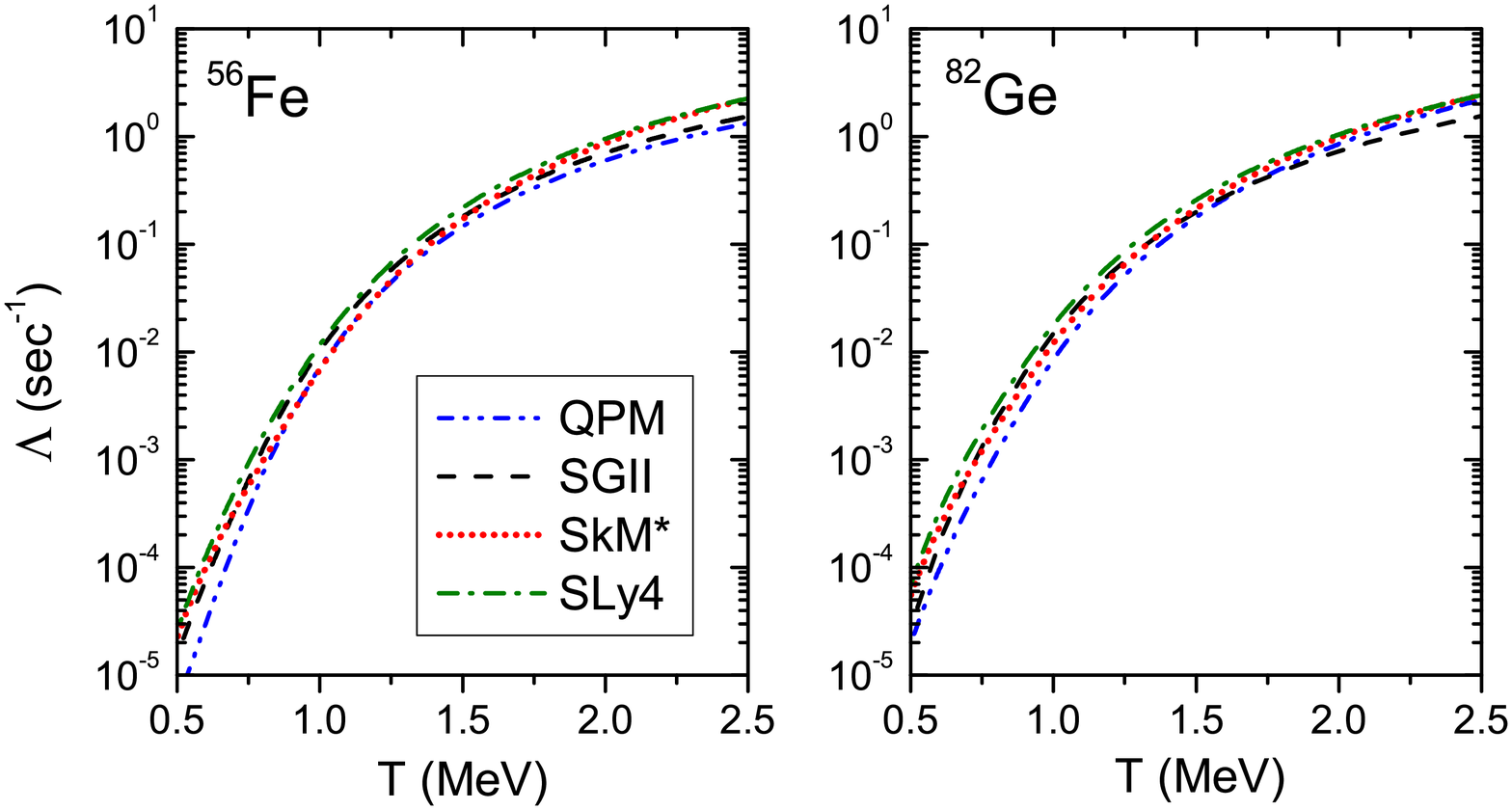}
\caption{(Color online) Neutrino-antineutrino pair emission rates for $^{56}$Fe and $^{82}$Ge as functions of temperature. }
 \label{emission_rates}
 \end{centering}
\end{figure}

In Fig.~\ref{emission_rates} we illustrate the temperature dependence  of the emission rates for $^{56}$Fe and $^{82}$Ge computed by using the Skyrme-TQRPA thermal strength
functions. For comparison, we also show the results from Ref.~\cite{Dzhioev_PAN77} obtained within the QPM-TQRPA calculations. As expected, the emission rate demonstrates a strong
thermal enhancement and from the above discussion it is clear that the main reason for that is the  thermal population and the subsequent decay of the GT$_0$ resonance.
The emission rates calculated with various residual forces differ by up to a factor five and the difference  is most significant at low temperatures.

\section{Conclusion}\label{conclusion}

Cross sections and rates for neutral-current neutrino-nucleus reactions have been calculated at supernova conditions for the two representative nuclei $^{56}$Fe and $^{82}$Ge. The thermal effects were treated within the Thermal Quasiparticle Random-Phase Approximation combined with Skyrme energy density functional theory.

The calculations with three different Skyrme functionals show the same thermal effects on the GT$_0$ strength function as those found in a previous study based on the QPM-TQRPA model. In particular, increasing the temperature shifts the GT$_0$ centroid to lower energies and makes possible low-energy GT$_0$ transitions. However, the Skyrme based calculations predict larger strengths of thermally unblocked low-energy transitions than those obtained within the QPM-TQRPA. Moreover, since the detailed balance principle is  fulfilled in the TQRPA, this difference leads to larger strengths for negative-energy GT$_0$ transitions.

The present calculations reveal the same thermal effects on the inelastic neutrino-nucleus scattering cross-section as the QPM-TQRPA model and the hybrid-model calculations. Namely, the reaction thresholds are removed at finite temperature and the cross sections for low-energy neutrinos are significantly enhanced. However, due to a larger strength of thermally unblocked  low- and negative-energy transitions, the low-energy cross-sections calculated with the Skyrme forces are larger than those obtained within the other approaches. It is interesting to note  that different Skyrme forces predict finite-temperature cross sections which do not differ significantly.

Considering neutrino-pair emission, it was found that, at low temperatures, the spectrum of emitted neutrinos is quite sensitive to details of the GT$_0$ strength distribution.
Moreover, the spectrum calculated without assuming the Brink hypothesis is shifted to somewhat lower energies.
It has been demonstrated that  temperature increase leads to a considerable enhancement of the emission rate due to thermal population of the GT$_0$ resonance.
For the temperatures considered, the absolute spread in the emission rates computed with different TQRPA models is less than an order of magnitude.

The present study can be considered as a first step toward a reliable description of supernova  neutrino-nucleus processes
  in a self-consistent microscopic approach. Based on the TFD formalism, the method allows us to evaluate the cross sections and rates in a thermodynamically
  consistent way, i.e.,  without assuming the Brink hypothesis and without violation the detailed-balance principle. The approach is not restricted by iron-group
  nuclei and can be applied for more massive neutron-rich nuclei where the shell-model diagonalization approach is not feasible.
  Further improvements and extensions of the current version of the model will be made along the lines discussed below.

  In our RPA calculations we explore a Landau-Migdal form of the residual interaction
   which does not include the spin-orbit term. Consequently, our calculations are not, strictly speaking, fully self-consistent. The spin-orbit term of the
  residual interaction has been shown to be very small in the case of charge-exchange GT excitations~\cite{Fracasso_PRC76}. Nevertheless, to  explicitly estimate the effect
  of the two-body spin-orbit residual interaction on the charge-neutral GT mode, fully self-consistent Skyrme - RPA calculations are needed.

   In the present study we consider Skyrme parametrizations  that do not include the tensor force. However, the tensor force may play an important role in both
   ground states and excited states properties~\cite{Sagawa_PPNP76} and,  therefore, needs to be taken into account.
  For charge-exchange GT excitations, the effects of tensor correlations have been analyzed in Ref.~\cite{Bai_PLB675} by using
  self-consistent  Hartree-Fock + RPA calculations with the Skyrme interaction SIII. It was found that tensor correlations are quite strong in that case and lead
  to a lowering of the main GT peak by about 2~MeV in $^{90}$Zr and $^{208}$Pb, which is accompanied by a shift of a sizable fraction of the energy weighted sum rule
  to the energy region above 30~MeV. If the same effects are present for charge-neutral GT excitations, the inclusion of tensor forces would effect the low-energy cross sections
  as well as the energy distribution of scattered and emitted neutrinos.

   Our present calculations are limited by the RPA level and do not include many-body correlations, which are responsible for the detailed fragmentation of the thermal strength function.
  The fragmentation is important at  low neutrino energies when  cross sections and rates are sensitive to the details of GT$_0$ strength distribution.
  Moreover, due the mixing of RPA states with more complex (e.g., 2p-2h) configurations, some part of the GT$_0$ strength may
  be shifted to higher energies, as was found for the charge-exchange GT strength~\cite{Bertsch_PRC26}. This effect should be important for the  neutrino pair production.
  Since the phase factor $E^5$ in Eq.~\eqref{decay_rate} favours the contribution from higher energies, the upward shift of the GT$_0$ strength can result
  in more energetic neutrinos  emitted at $T\gtrsim1$~MeV, than it is shown in Figs.~\ref{partial_rates} and~\ref{E_mean}.
   In cold nuclei the inclusion of higher-order correlations beyond the RPA level was considered within the QPM~[16] by coupling of one-phonon states  with two-phonon configurations.
  By exploiting a separable approximation for the Skyrme interaction one could consider phonon coupling at finite temperature within a self-consistent theory.
  This is planned for  future work.

\begin{acknowledgments}

This work was supported by the Heisenberg-Landau Program.

\end{acknowledgments}

\appendix*
\section{}\label{app}

Within the TQRPA thermal phonons are defined as a linear superposition of creation and annihilation operators of thermal quasiparticle pairs
\begin{multline}\label{phonon}
  Q^\dag_{J M i}=\frac12\sum_{\tau=n,p}{\sum_{j_1j_2}}^\tau
 \Bigl\{\psi^{Ji}_{j_1j_2}[\beta^\dag_{j_1}\beta^\dag_{j_2}]^J_M +
 \widetilde\psi^{J i}_{j_1j_2}[\widetilde\beta^\dag_{\overline{\jmath_1}}
 \widetilde\beta^\dag_{\overline{\jmath_2}}]^J_M
 \\ +
 i\eta^{J i}_{j_1j_2}[\beta^\dag_{j_1}
  \widetilde\beta^\dag_{\overline{\jmath_2}}]^J_M+
 i\widetilde\eta^{J i}_{j_1j_2}[\beta^\dag_{\overline{\jmath_1}}
  \widetilde\beta^\dag_{j_2}]^J_M
+
 \phi^{J i}_{j_1j_2}[\beta_{\overline{\jmath_1}}\beta_{\overline{\jmath_2}}]^J_M  \\
+ \widetilde\phi^{J i}_{j_1j_2}[\widetilde\beta_{j_1}
 \widetilde\beta_{j_2}]^J_M +
  i\xi^{J i}_{j_1j_2}[\beta_{\overline{\jmath_1}}
  \widetilde\beta_{j_2}]^J_M
+i\widetilde\xi^{J i}_{j_1j_2}[\beta_{j_1}
  \widetilde\beta_{\overline{\jmath_2}}]^J_M\Bigr\}.
\end{multline}
To clarify the physical meaning of different terms in~\eqref{phonon}, we note that the creation of a negative energy tilde thermal quasiparticle corresponds to the annihilation of a thermally excited Bogoliubov quasiparticle or, which is the same, to the creation of a quasihole state (see~\cite{Dzhioev_PRC81} for more details). Therefore, at finite temperature, single-particle transitions involve excitations of three types: (1) two-quasiparticle excitations described by the operator $\beta^\dag_{j_1}\beta^\dag_{j_2}$ and having energy $\varepsilon^{(+)}_{j_1j_2}=\varepsilon_{j_1}+\varepsilon_{j_2}$, (2) one-quasiparticle one-quasihole excitations
described by the operator $\beta^\dag_{j_1}\widetilde\beta^\dag_{j_2}$ and having energy $\varepsilon^{(-)}_{j_1j_2}=\varepsilon_{j_1}-\varepsilon_{j_2}$, and (3) two-quasihole excitations described by the operator $\widetilde\beta^\dag_{j_1}\widetilde\beta^\dag_{j_2}$ and having energy
$-\varepsilon^{(+)}_{j_1j_2}$. The last two types are possible only at $T\ne0$. Therefore, due to single-particle
transitions involving the annihilation of thermally excited Bogoliubov quasiparticles, the spectrum of thermal phonons contains negative- and low-energy states that do not exist at zero temperature. These "new" phonon states can be interpreted as thermally unblocked transitions between nuclear excited states.

In what follows we consider phonons of unnatural parity $\pi = -(-1)^J$. They are generated by the $L=J\pm1$ components of the residual interaction~\eqref{V}. To find the energy and the structure of the thermal phonons we apply the equation-of-motion method
\begin{equation}\label{eq_mot}
\langle|\delta Q,[{\cal H},Q^\dag]]|\rangle = \omega(T)\langle|[\delta Q,Q^\dag]|\rangle
\end{equation}
under two additional constraints: (a) the phonon operators obey Bose commutation relations, and (b) the phonon vacuum obeys the thermal  state condition~\eqref{TSC}. The first constraint leads to the normalization condition for the phonon amplitudes
 \begin{align}\label{constraint}
\frac12 \sum_{\tau=n,p}{\sum_{j_1 j_2}}^\tau\Bigl(&
  \psi^{Ji}_{j_1j_2}\psi^{J i'}_{j_1j_2}+
 \widetilde\psi^{J i}_{j_1j_2}\widetilde\psi^{J i'}_{j_1j_2}+
  \eta^{J i}_{j_1j_2}\eta^{J i'}_{j_1j_2}
  \notag\\
  &+\widetilde\eta^{J i}_{j_1j_2}\widetilde\eta^{Ji'}_{j_1j_2}
   -\phi^{J i}_{j_1j_2}\phi^{J i'}_{j_1j_2}-
  \widetilde\phi^{J i}_{j_1j_2}\widetilde\phi^{Ji'}_{j_1j_2}
  \notag\\
  &-\xi^{J i}_{j_1j_2} \xi^{\lambda i'}_{j_1j_2}
  - \widetilde\xi^{J i}_{j_1j_2}\widetilde\xi^{J i'}_{j_1j_2}\Bigr)=\delta_{ii'},
      \end{align}
while the last assumption yields the following relations between amplitudes
\begin{align}\label{constraint1}
  \binom{\widetilde\psi}{\widetilde\phi}^{J i}_{j_1j_2}&=\frac{y_{j_1}y_{j_2}-\mathrm{e}^{-\omega_{Ji}/2T}x_{j_1}x_{j_2}}{\mathrm{e}^{-\omega_{Ji}/2T}y_{j_1}y_{j_2}-x_{j_1}x_{j_2}}
\binom{\phi}{\psi}^{J i}_{j_pj_n},
\notag \\
 \binom{\widetilde\eta}{\widetilde\xi}^{J i}_{j_1j_2}&=\frac{y_{j_1}x_{j_2}-\mathrm{e}^{-\omega_{Ji}/2T}x_{j_1}y_{j_2}}{\mathrm{e}^{-\omega_{Ji}/2T}y_{j_1}x_{j_2}-x_{j_1}y_{j_2}}
\binom{\xi}{\eta}^{J i}_{j_1j_2}.
\end{align}
Here, $x_j$ and $y_j$ ($x^2_j+y^2_j=1$) are the coefficients of the so-called thermal transformation which establishes a connection between Bogoliubov and thermal quasiparticles. Note that $y_j$ are given by the nucleon Fermi-Dirac function and they define a number of thermally excited Bogoliubov quasiparticles in the thermal vacuum~(see Ref.~\cite{Dzhioev_PRC81} for more details).

To derive the TQRPA equations it is convenient to introduce the following linear combinations of amplitudes
 \begin{align}\label{gw}
 \binom{g}{w}^{J i}_{j_1j_2}&=
 \psi^{Ji}_{j_1j_2}\pm\phi^{Ji}_{j_1j_2},
 & \binom{\widetilde g}{\widetilde w}^{J i}_{j_1j_2}&=
 \widetilde\psi^{J i}_{j_1j_2}\pm\widetilde\phi^{Ji}_{j_1j_2},
 \notag\\[3mm]
\binom{t}{s}^{Ji}_{j_1j_2}&=
 \eta^{J i}_{j_1j_2}\pm\xi^{J i}_{j_1j_2},
 & \binom{\widetilde t}{\widetilde s}^{J i}_{j_1j_2}&=
 \widetilde\eta^{J i}_{j_1j_2}\pm\widetilde \xi^{J i}_{j_1j_2}.
\end{align}
Then, from Eq.~\eqref{constraint1}, it follows that
\begin{align}\label{gwgw}
  \binom{g}{w}^{J i}_{j_1j_2}&= (x_{j_1}x_{j_2}-\mathrm{e}^{-\omega_{Ji}/2T}y_{j_1}y_{j_2})\binom{G}{W}^{J i}_{j_1j_2}
  \notag\\
  \binom{\widetilde g}{\widetilde w}^{J i}_{j_1j_2}&= \mp(y_{j_1}y_{j_2}-\mathrm{e}^{-\omega_{Ji}/2T}x_{j_1}x_{j_2})\binom{G}{W}^{J i}_{j_1j_2}
  \notag\\
  \binom{t}{s}^{Ji}_{j_1j_2}&=(x_{j_1}y_{j_2}-\mathrm{e}^{-\omega_{Ji}/2T}y_{j_1}x_{j_2})\binom{T}{S}^{J i}_{j_1j_2}
  \notag\\
  \binom{\widetilde t}{\widetilde s}^{Ji}_{j_1j_2}&=\mp(y_{j_1}x_{j_2}-\mathrm{e}^{-\omega_{Ji}/2T}x_{j_1}y_{j_2})\binom{T}{S}^{J i}_{j_1j_2},
\end{align}
where $G,~W,~T$ and $S$ are normalized according to
\begin{align}\label{norm}
  \frac12&\sum_{\tau=p,n}{\sum_{j_1j_2}}^\tau
   \Bigl(G^{Ji}_{j_1j_2}W^{Ji'}_{j_1j_2}(1-y^2_{j_1}-y^2_{j_2})
  \notag\\
  &~~~- T^{Ji}_{j_1j_2}S^{Ji'}_{j_1j_2}(y^2_{j_1}-y^2_{j_2})\Bigr)
  =\delta_{ii'}/(1-\mathrm{e}^{-\omega_{Ji}/T}).
\end{align}

From the equation of motion~\eqref{eq_mot} we get the system of TQRPA equations for the unknown $G,~W,~T$, and $S$ variables and phonon energies
\begin{align}\label{eqTQRPA}
  &G^{Ji}_{j_1j_2}=\frac{\omega_{Ji}}{\varepsilon^{(+)}_{j_1j_2}}W^{Ji}_{j_1j_2},
  \notag\\
  &T^{Ji}_{j_1j_2}=\frac{\omega_{Ji}}{\varepsilon^{(-)}_{j_1j_2}}S^{Ji}_{j_1j_2},
  \notag\\
  &[\varepsilon^{(+)}_{j_1j_2}]^2W^{Ji}_{j_1j_2} -\frac{\varepsilon^{(+)}_{j_1j_2}}{\hat J^2}\sum_{k=1}^N\sum_{L=J\mp1} g^{(LJk)}_{j_1j_2} u^{(-)}_{j_1j_2}\times
  \notag\\
  &\Bigl\{(\kappa^{(k)}_0 + \kappa^{(k)}_1) D^{i}_{LJk} (\tau) + (\kappa^{(k)}_0 - \kappa^{(k)}_1) D^{i}_{LJk}(-\tau)\Bigl\}
  \notag\\
  &~~~~~=\omega^2_{Ji} W^{Ji}_{j_1j_2},
  \notag\\
  &[\varepsilon^{(-)}_{j_1j_2}]^2S^{Ji}_{j_1j_2} -\frac{\varepsilon^{(-)}_{j_1j_2}}{\hat J^2}\sum_{k=1}^N\sum_{L=J\mp1} g^{(LJk)}_{j_1j_2} v^{(+)}_{j_1j_2}\times
  \notag\\
  &\Bigl\{(\kappa^{(k)}_0 + \kappa^{(k)}_1) D^{i}_{LJk}(\tau) + (\kappa^{(k)}_0 - \kappa^{(k)}_1) D^{i}_{LJk}(-\tau)\Bigl\}
  \notag\\
  &~~~~~=\omega^2_{Ji} S^{Ji}_{j_1j_2},
\end{align}
where
\begin{align*}
D^{i}_{LJk}(\tau) =& {\sum_{j_1j_2}}^\tau  g^{(LJk)}_{j_1j_2}\Bigl\{ u^{(-)}_{j_1j_2}(1-y^2_{j_1}-y^2_{j_2})W^{Ji}_{j_1j_2}-
   \notag\\
               &v^{(+)}_{j_1j_2} (y^2_{j_1}-y^2_{j_2})S^{Ji}_{j_1j_2}\Bigr\}.
\end{align*}
In the above equations we have introduced the following combinations of the Bogoliubov ($u,v$)-coefficients: $u^{(-)}_{j_1j_2}=u_{j_1}v_{j_2}-v_{j_1}u_{j_2}$, $v^{(+)}_{j_1j_2}=u_{j_1}u_{j_2}+v_{j_1}v_{j_2}$.

Because of the separable form of the residual interaction the TQRPA equations can be reduced to the set of equations for $D^{i}_{LJk}$
\begin{equation}
  \left(
    \begin{array}{cc}
      \mathcal{M}_{J-1,J-1}-1 & \mathcal{M}_{J-1,J+1} \\
      \mathcal{M}_{J+1,J-1}   & \mathcal{M}_{J+1,J+1}-1 \\
    \end{array}
    \right)
    \left(
      \begin{array}{c}
        \mathcal{D}_{J-1,J} \\
        \mathcal{D}_{J+1,J}\\
      \end{array}
  \right)=0.
\end{equation}
Here $\mathcal{M}_{L,L'}$ is the $2N\times2N$ matrix
\begin{widetext}
\begin{equation}
  \mathcal{M}^{kk'}_{LL'} =
  \left(
    \begin{array}{cc}
      (\kappa_0^{(k')}+ \kappa_1^{(k')})\mathcal{X}^{kk'}_{LL'}(p)& (\kappa_0^{(k')}- \kappa_1^{(k')})\mathcal{X}^{kk'}_{LL'}(p) \\
      (\kappa_0^{(k')}- \kappa_1^{(k')})\mathcal{X}^{kk'}_{LL'}(n) & (\kappa_0^{(k')}+ \kappa_1^{(k')})\mathcal{X}^{kk'}_{LL'}(p) \\
    \end{array}
  \right),~~~~1\le k,k'\le N.
\end{equation}
\end{widetext}
with the following matrix elements $\mathcal{X}^{kk'}_{LL'}(\tau)$:
\begin{align}\label{MatEl}
  \mathcal{X}^{kk'}_{LL'}(\tau)=&\frac{1}{\hat J^2}{\sum_{j_1j_2}}^\tau g^{(LJk)}_{j_1j_2}g^{(L'Jk')}_{j_1j_2}\Biggl\{
  \frac{\varepsilon^{(+)}_{j_1j_2}[u^{(-)}_{j_1j_2}]^2}{ [\varepsilon^{(+)}_{j_1j_2}]^2-\omega^2}
  \notag\\
  &\times(1-y^2_{j_1}-y^2_{j_2}) -
  \frac{\varepsilon^{(-)}_{j_1j_2}[v^{(+)}_{j_1j_2}]^2}{ [\varepsilon^{(-)}_{j_1j_2}]^2-\omega^2}(y^2_{j_1}-y^2_{j_2})\Biggr\}.
\end{align}
The $2N$-vector $\mathcal{D}_{LJ}$ has the components
\begin{equation*}
  \mathcal{D}_{LJk}=\left(
                     \begin{array}{c}
                       D_{LJk}(p) \\
                       D_{LJk}(n) \\
                     \end{array}
                   \right), ~~~1\le k \le N.
\end{equation*}

Thus, the TQRPA eigenvalues $\omega_{Ji}$ are the roots of the secular equation
\begin{equation}
 \mathrm{det}
  \left(
    \begin{array}{cc}
      \mathcal{M}_{J-1,J-1}-1 & \mathcal{M}_{J-1,J+1} \\
      \mathcal{M}_{J+1,J-1}   & \mathcal{M}_{J+1,J+1}-1 \\
    \end{array}
    \right)=0.
\end{equation}
while the phonon amplitudes corresponding to the TQRPA eigenvalue $\omega_{Ji}$ are determined by Eqs.~\eqref{gw}, \eqref{gwgw}, and \eqref{eqTQRPA}, taking into account the normalization condition~\eqref{norm}.



%

\end{document}